\documentclass[graybox]{svmult}
\usepackage{mathptmx}
\usepackage{makeidx} 
\usepackage[utf8]{inputenc}
\usepackage{amsmath}
\usepackage{amsfonts}
\usepackage{amssymb}
\usepackage[margin=1in]{geometry}
\usepackage{graphicx}
\usepackage{hyperref}
\usepackage[hyphenbreaks]{breakurl}
\usepackage{epstopdf}
\usepackage{algorithm}
\usepackage{algpseudocode}
\usepackage{hyperref}
\usepackage{bm}
\usepackage{caption}
\usepackage[bottom]{footmisc}
\usepackage{listings}
\usepackage{color}
\usepackage{multicol}

\definecolor{codegreen}{rgb}{0,0.6,0}
\definecolor{codegray}{rgb}{0.5,0.5,0.5}
\definecolor{codepurple}{rgb}{0.58,0,0.82}
\definecolor{backcolour}{rgb}{0.95,0.95,0.92}
 
\lstdefinestyle{mystyle}{
    backgroundcolor=\color{backcolour},   
    commentstyle=\color{codegreen},
    keywordstyle=\color{magenta},
    numberstyle=\tiny\color{codegray},
    stringstyle=\color{codepurple},
    basicstyle=\footnotesize,
    breakatwhitespace=false,         
    breaklines=true,                 
    captionpos=b,                    
    keepspaces=true,                 
    numbers=left,                    
    numbersep=5pt,                  
    showspaces=false,                
    showstringspaces=false,
    showtabs=false,                  
    tabsize=2
}
 
\begin{document}
\title*{Multi-omic Network Regression: Methodology, Tool and Case Study}
\titlerunning{Multi-omic Network Regression}
\author{Vandan Parmar, Pietro Li{\'{o}}}
\institute{Vandan Parmar \at University of Cambridge, \email{vandan.parmar@googlemail.com}
\and Pietro Li{\'{o}} \at University of Cambridge, \email{pl219@cam.ac.uk}}
\maketitle

\abstract*{
The analysis of biological networks is characterized by the definition of precise linear constraints used to cumulatively reduce the solution space of the computed states of a multi-omic (for instance metabolic, transcriptomic and proteomic) model. In this paper, we attempt, for the first time, to combine metabolic modelling and networked Cox regression, using the metabolic model of the bacterium  \textit{Helicobacter Pylori}. This enables a platform both for quantitative analysis of networked regression, but also testing the findings from network regression (a list of significant vectors and their networked relationships) on \textit{in vivo} transcriptomic data. Data generated from the model, using flux balance analysis to construct a Pareto front, specifically, a trade-off of Oxygen exchange and growth rate and a trade-off of Carbon Dioxide exchange and growth rate, is analysed and then the model is used to quantify the success of the analysis. It was found that using the analysis, reconstruction of the initial data was considerably more successful than a pure noise alternative. Our methodological approach is quite general and it could be of interest for the wider community of complex networks researchers; it is implemented in a software tool, MoNeRe, which is freely available through the Github platform. 
}
\abstract{
The analysis of biological networks is characterized by the definition of precise linear constraints used to cumulatively reduce the solution space of the computed states of a multi-omic (for instance metabolic, transcriptomic and proteomic) model. In this paper, we attempt, for the first time, to combine metabolic modelling and networked Cox regression, using the metabolic model of the bacterium  \textit{Helicobacter Pylori}. This enables a platform both for quantitative analysis of networked regression, but also testing the findings from network regression (a list of significant vectors and their networked relationships) on \textit{in vivo} transcriptomic data. Data generated from the model, using flux balance analysis to construct a Pareto front, specifically, a trade-off of Oxygen exchange and growth rate and a trade-off of Carbon Dioxide exchange and growth rate, is analysed and then the model is used to quantify the success of the analysis. It was found that using the analysis, reconstruction of the initial data was considerably more successful than a pure noise alternative. Our methodological approach is quite general and it could be of interest for the wider community of complex networks researchers; it is implemented in a software tool, MoNeRe, which is freely available through the Github platform. 
}

\section{Introduction}

Large scale networks have increasingly become ubiquitous with the development of the Internet of Things \cite{Atzori2010}, the smart grid \cite{Amin2005,Farhangi2010} and smart motorways \cite{Chien1997}. These networks have also been developed for biological purposes, both to identify significant genes for the growth of cancer (networked Cox regression \cite{Cox1972}) and model whole cell behaviour (metabolic modelling \cite{Orth2010}).

Both networked Cox regression and multi-omic metabolic modelling are new techniques which aim to develop the more established Cox regression and metabolic modelling. Cox regression attempts to find significant genes within a dataset of gene expression data. Previously this would have only been found by taking samples from patients or bacteria. However, with multi-omic metabolic modelling, gene expressions can be incorporated into the modelling framework, meaning gene expression data can be generated computationally using these models. Where normally it is difficult to verify the effectiveness of Cox regression, when the gene expression data is created using a computational model, the same computational model can be used to test the predictions of the Cox regression.

This paper focuses on the metabolic network of \textit{Helicobacter Pylori}, a bacterium usually found in the stomach. Originally found in 1982 to be causing stomach ulcers \cite{Marshall1984} and linked to stomach cancer \cite{Forman1991}, \textit{Helicobacter Pylori} causes no symptoms in approximately 80\% of infected individuals and is thought to be important for the stomach ecology \cite{Blaser2006}. 

\twocolumn

\begin{figure}
\centering
\includegraphics[scale=0.65]{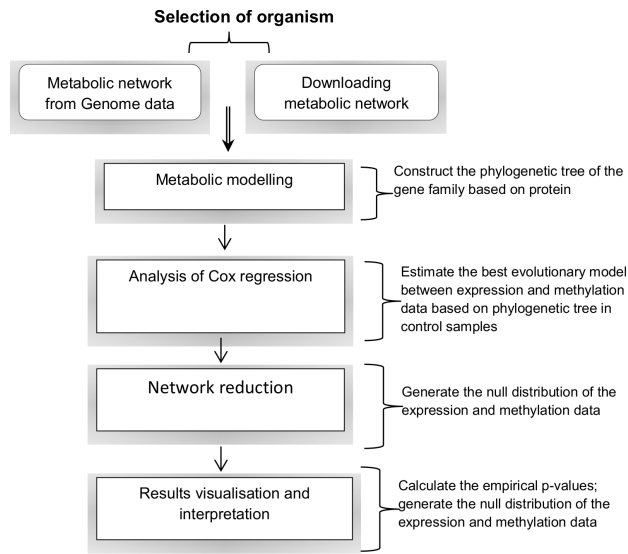}
\caption{A visual representation of the work flow developed in this project.}
\label{fig:scheme}
\end{figure}

Datasets were created using the \textit{H. Pylori} metabolic model, the regression was analysed and tuned, and the success of the overall regression was then tested by attempting to reconstruct the original data. A visual representation of the developed work flow is shown in figure \ref{fig:scheme}. The following section contains a summary of related research, section \ref{sec:theory} details the theory and computational methods required, results and discussion is provided in section \ref{sec: res and disc} and concluding remarks are given in section \ref{sec:conc}. The code from this paper can be found at: \url{https://github.com/vandanparmar/MoNeRe}.

\section{Background}
\subsection{Metabolic Modelling}
The emergence of new sequencing methods allowing for complete organism genomes to be sequenced and the explosion of computing power over the past 30 years, has lead to the emergence of Systems Biology as a field of study \cite{LeNovere2007}. This has allowed increasingly good models of a cell to be developed. Specifically, models of a cell's metabolism (the set of chemical reactions taking place in the cell) can be made. The first was created by Tomita et al. \cite{Tomita1999}, however more recent metabolic models are designed to include data from other sources. These include the proteome (set of proteins expressed by a genome) and the transcriptome (set of messenger RNA molecules expressed by the genome) and are collectively known as `omic' data. A multi-omic model of this type can be predictive  \cite{Angione2015}, often aiming to find unexpected behaviour which can then be tested \textit{in vitro}.

Once a model is constructed, analysis is primarily carried out using Flux Balance Analysis (FBA). There are several existing toolboxes which provide frameworks for carrying out FBA on genome scale metabolic models, including COBRA \cite{Ebrahim2013}, METRADE \cite{Angione2015} and DFBALab \cite{AlbertoGomezPaulBarton}, in this study, COBRA will be used for FBA calculations. FBA is a particularly powerful method, as it reduces the multidimensional search for an optimal metabotype (set of fluxes through all reactions in the system) to a linear program \cite{Sierksma2002}, under the assumptions of a steady state.  FBA has been used to model bacteria such as \textit{Escheria Coli} \cite{Angione2015c} and more recently breast cancer cells \cite{Angione2017a}.  

\subsection{Cox Regression}
Cox regression \cite{Cox1972} was created as an extension of Kaplan-Meier survival analysis \cite{Kaplan1958}. Kaplan-Meier survival analysis aims to find the probability of survival for a species whilst incorporating data from living individuals. The Cox model furthers this by including extra data (other than survival time) about each individual. This data could be anything, but is most often gene expression data. More recently, this has been extended by adding network constraints \cite{Iuliano2016} both to enhance the regression and discover networks relating to the regression. This has been used successfully identify significant biomarkers in predicting the outcome of Ovarian cancer treatment \cite{Zhang2013}.

\section{Theory and Computational Methods}
\label{sec:theory}

\subsection{Network Regression}
\label{ssec: Network Regression}
Network regression \cite{Li2010} is the primary method of Pareto data analysis used in this study. Compared to linear regression, networked constraints are imposed on regression coefficients, enforcing similarity between the regression coefficients of linked genes.

Cox regression \cite{Cox1972} is used as a basis for the networked regression used. The Cox hazard model assumes a hazard function, $h(t|\textbf{X}_i)$, the risk of death at time $t$,

\begin{equation}
h(t|\textbf{X}_i) = h_0(t)\exp(\textbf{X}_i \cdot \bm{\beta})
\end{equation}

where $\textbf{X}_i$ is the $i$th gene expression profile, $h_0(t)$ is the baseline hazard and $\bm{\beta}$ is the vector of regression parameters.

Within the Cox model, the regression parameters are estimated by maximizing the Cox log-partial likelihood,
\begin{equation}
p_l( \bm{ \beta } ) = \sum_{i=1}^{n} \delta_i \left\lbrace \textbf{X}_i \cdot \bm{\beta} - \log \left[ \sum_{j \in \mathcal{R}(t_i)} \exp( \textbf{X}_j \cdot \bm{\beta} ) \right] \right\rbrace
\end{equation}

where $\delta_i =0$ if the $i$th individual is alive and $\delta_i = 1$ if the $i$th individual is dead and $\mathcal{R}(t_i)$ is the set of individuals alive at time $t_i$.

The network constraints can be added in a variety of ways \cite{Iuliano2016}, in this study a networked penalty, $p_n(\bm{\beta})$, was subtracted from the log-partial likelihood,
\begin{equation}
p_n(\bm{\beta}) = - \lambda \alpha |\bm{\beta}|^2 - \lambda (1 - \alpha) \bm{\beta}^T \textbf{S} \bm{\beta}
\end{equation}
 
where $\lambda$ and $\alpha$ are tuning parameters, $\textbf{I}$ is the identity matrix and $\textbf{S}$ is a weight matrix related to the network. The second term is the network penalty itself and the first term is a sparsity constraint, forcing the regression to identify only a few significant genes. $\alpha$ controls the ratio between the sparsity and network constraints and $\lambda$ controls the ratio between the simple Cox regression and the networked constraints.

$\textbf{S}$ is computed by first computing $\textbf{W}$, a gene coexpression network,
\begin{align}
w_{ij} &= \text{PMCC}(g_i,g_j) \\
(\textbf{W})_{ij} &= \left\lbrace \begin{aligned}
 w_{ij}^2 \quad & w_{ij}^2 > \zeta \\
 0  \quad & w_{ij}^2 < \zeta 
\end{aligned} \right.
\end{align}

where PMCC indicates the product moment correlation coefficient, so the values of $\textbf{W}$ are between $\zeta$ and 1. The value of $\zeta$ can be tuned to change the connectivity of the network. The adjacency matrix of the network is thus the ceiling of $\textbf{W}$. From this $\textbf{S}$ can be calculated,

\begin{equation}
(\textbf{S})_{ij} = \left\lbrace
\begin{aligned}
1& \quad i=j \, \text{and} \, d_i \neq0 \\
\frac{-(\textbf{W})_{ij}}{\sqrt{d_i d_j}}& \quad (i,j) \in \mathcal{E} \\
0& \quad \text{otherwise}
\end{aligned}
\right.
\label{eq:s_description}
\end{equation}

where $d_i$ is the degree of the $i$th node and $\mathcal{E}$ is the set of edges within the network. $\textbf{S}$ is a weighted version of the normalized Laplacian of a graph \cite{Weisstein} and this matrix is always positive semidefinite (i.e. $\bm{\beta}^T \textbf{S} \bm{\beta} \geq 0$).

As $\textbf{S}$ is always positive semidefinite, all terms in the penalty function $p_l(\bm{\beta}) - p_n(\bm{\beta})$ are convex (surprisingly $\log \sum_i \exp(\textbf{X}_i \cdot \bm{\beta})$ is convex \cite{Boyd2004}), meaning that convex optimization \cite{Diamond2016} can be used to perform the regression.

In the case of a Pareto front, we perform regression on each side of the phase transition individually, considering individuals in one phase to be alive and those in the other dead. The optimization can then be simply formed as,

\begin{equation}
\begin{aligned}
	&\max  \left\lbrace p_l(\bm{\beta}) - p_n(\bm{\beta}; \lambda, \alpha) \right\rbrace \\
&\text{s. t.} \qquad  |\bm{\beta}| \leq 1
\end{aligned}
\end{equation}

where $x_p$ is the value of an objective function at the phase transition and $x_j$ is the corresponding objective value of the individual.
\subsection{Network Reduction}
\label{ssec:network reduction}
\begin{figure}[H]
	\centering
	\includegraphics[scale=0.32]{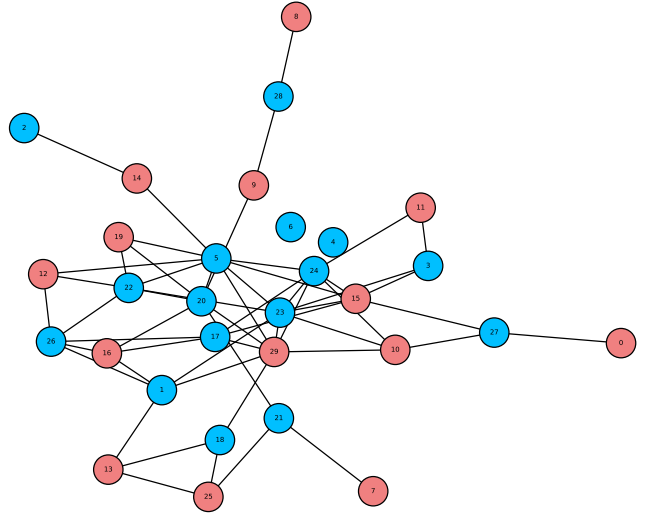}
	\caption{A network showing nodes included in the reduced network in blue and others in red. Red nodes are connected to at least one blue node.}
	\label{fig:reduced_network}
\end{figure}

When reconstructing Pareto fronts after completing regression, to further impose network constraints, a subsection of the relevant network containing significant nodes can be found. This was done by creating a subset of the network where each node is at a maximum of one edge from a node in the subset as shown in figure \ref{fig:reduced_network}. The procedure for this is as follows,

\begin{enumerate}
	\item Create a new copy of the graph (to edit)
	\item Create a list of nodes, sorted by degree descending. 
	\item Remove the first node in the sorted list, add it to the reduced network, add its neighbours to a list of accounted for nodes, remove the node from the graph.
	\item Update the sorted list of nodes.
	\item Repeat steps 2 and 3 until all nodes are accounted for.
\end{enumerate}

\subsection{Implementation Outline}
The code for the project was exclusively written in Python3, however, multiple libraries were used in conjunction. The external libraries used were: Numpy, Networkx, CVXpy, COBRA, DEAP, TQDM. Networkx is used for construction and calculation on networks, CVXpy for the network regression, COBRA for handling the metabolic model, DEAP for the genetic algorithm and TQDM for providing progress bars for code with longer run times. The \textit{H. Pylori} model \cite{Thiele2005} used is taken from BiGG Models \cite{King2016b}.

The code for this paper is available at: \linebreak \url{github.com/vandanparmar/MoNeRe}.

\section{Results and Discussion}
\label{sec: res and disc}

\subsection{Data Generation}
\begin{figure}[h]
	\centering
	\includegraphics[scale=0.5]{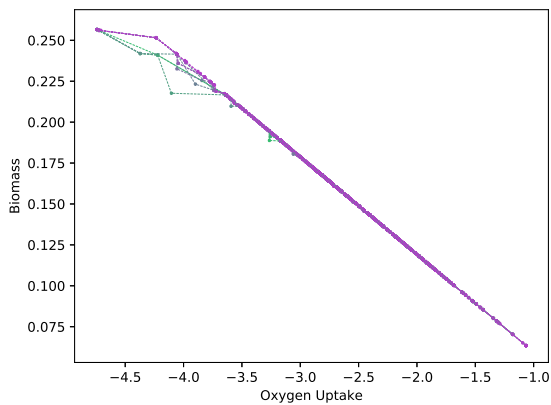}
	\caption{Graph showing the progression of the genetic algorithm, forming the Pareto front between Oxygen uptake and Biomass (cell growth rate). Points from earlier generations are in green, points from later generations are in purple. The general progression is to the upper right quartile, the non dominated area.}
	\label{fig:basic pareto}
\end{figure}

The network of focus in this study is the metabolic network of \textit{H. Pylori}. To see the way the genetic algorithm works, the Pareto front for the trade-off between Biomass (the amount of cell growth) and Oxygen uptake was found, shown in figure \ref{fig:basic pareto}. After exploring several trade-offs, this trade-off was chosen as it gives a particularly nice phase transition. The Oxygen, biomass trade-off is also closely linked to energy, thus we thought it likely that many genes would be involved. The values for Oxygen uptake are negative, thus maximising this has the effect of attempting to reduce the amount of Oxygen absorbed by the cell. This graph therefore shows the comparison of growth rate in aerobic compared to anaerobic conditions. We thus might expect that there would be a phase transition between the aerobic and anaerobic regime.

\begin{figure}[h]
	\centering
	\includegraphics[scale=0.5]{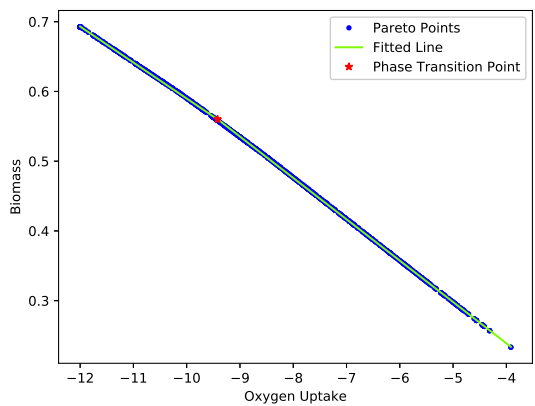}
	\caption{Graph showing a Pareto front with more points than in figure \ref{fig:basic pareto}. This Pareto front shows a phase transition, the gradient of the Pareto front changes from -0.051 to -0.059 on either side of the red star indicating the phase transition point.}
	\label{fig:dense pareto}
\end{figure}

The Pareto front was calculated again, with both more individuals in each generation and more generations, as shown in figure \ref{fig:dense pareto}. This shows a phase transition, with the Pareto front changing gradient from -0.051 to -0.059 either side of the transition. Transitions like these were found with a range of different objective functions. In order to investigate these, a simplified model was created.

\subsection{Loopless FBA}

It is clear that the initial flux bounds used within a model are significant. Changing flux bounds is equivalent to changing the environment (both internal and external) that a cell is operating in. In the literature, FBA is typically used to investigate a specific function or behaviour only, appropriate flux bounds can thus be found from experimentation or to enforce desired conditions \cite{Orth2010}. However, in this case, we are looking at arbitrary pairs of objectives, where the large flux bounds used in the \textit{H. Pylori} BiGG model \cite{King2016b} are not appropriate.

To find reasonable upper and lower bounds for each reaction, we use a technique known as Loopless FBA. This adds an additional thermodynamical constraint to FBA, which makes loops or cycles infeasible. Loopless FBA (implemented in the COBRApy toolbox \cite{Ebrahim2013})  was used to maximize and minimize the flux through each cell reaction. The maximum and minimum flux through each reaction across all of these runs was then used as the initial flux bound. In this way we avoid the need for biological data to set specific reaction constraints, though in practice this would still be ideal.

\subsubsection{Parameter Comparison}

The Cox based network regression, as described in section \ref{ssec: Network Regression}, has 3 parameters that can be tuned. $\lambda$ controls the ratio of the network constraints to the simple Cox regression, $\alpha$ controls the ratio between the network constraint and the sparsity constraint and $\zeta$ controls the connectivity of the gene coexpression network. These values can be found using cross-validation.

\begin{figure}[h]
	\centering
	\includegraphics[scale=0.44]{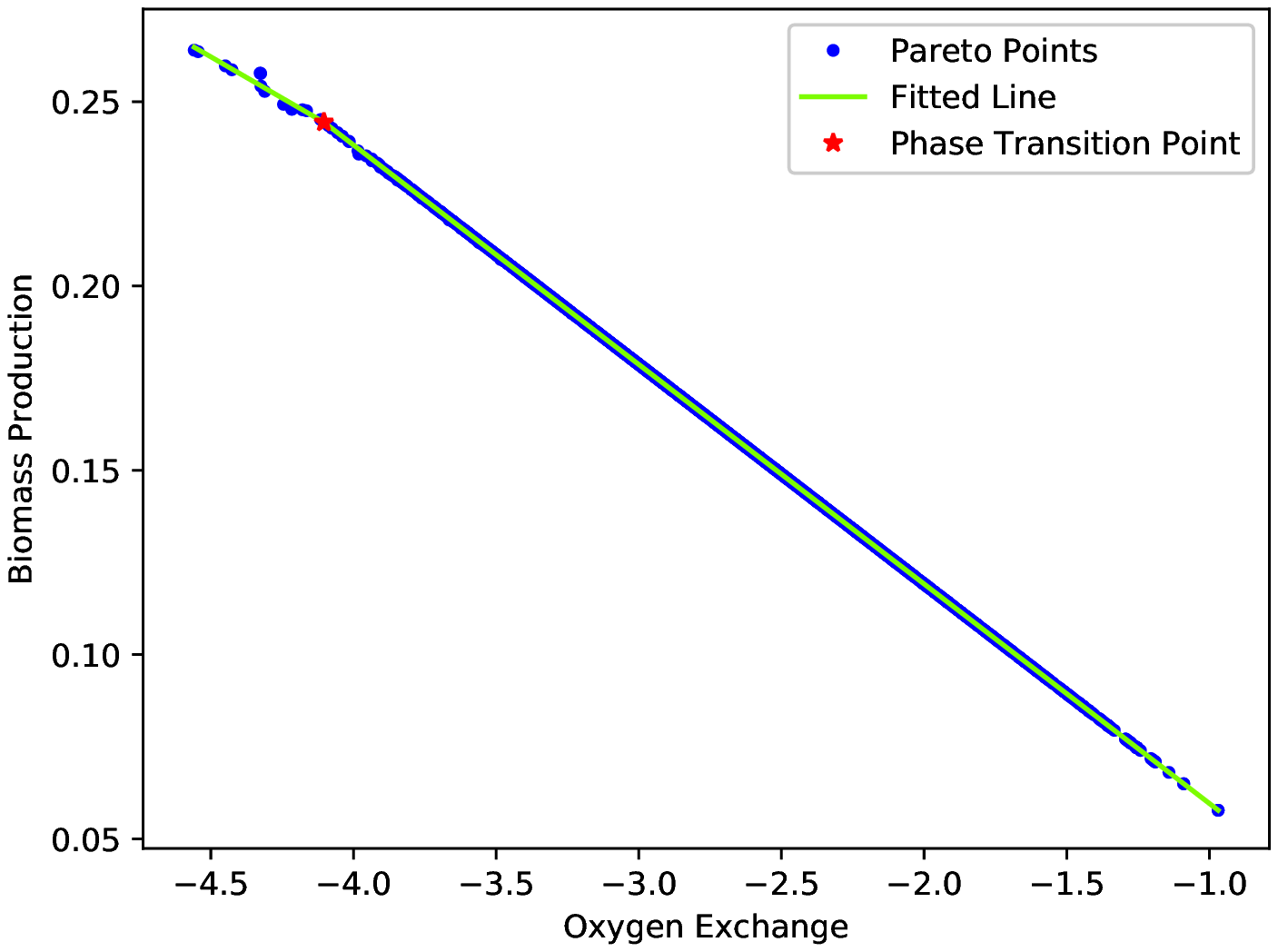}
	\includegraphics[scale=0.44]{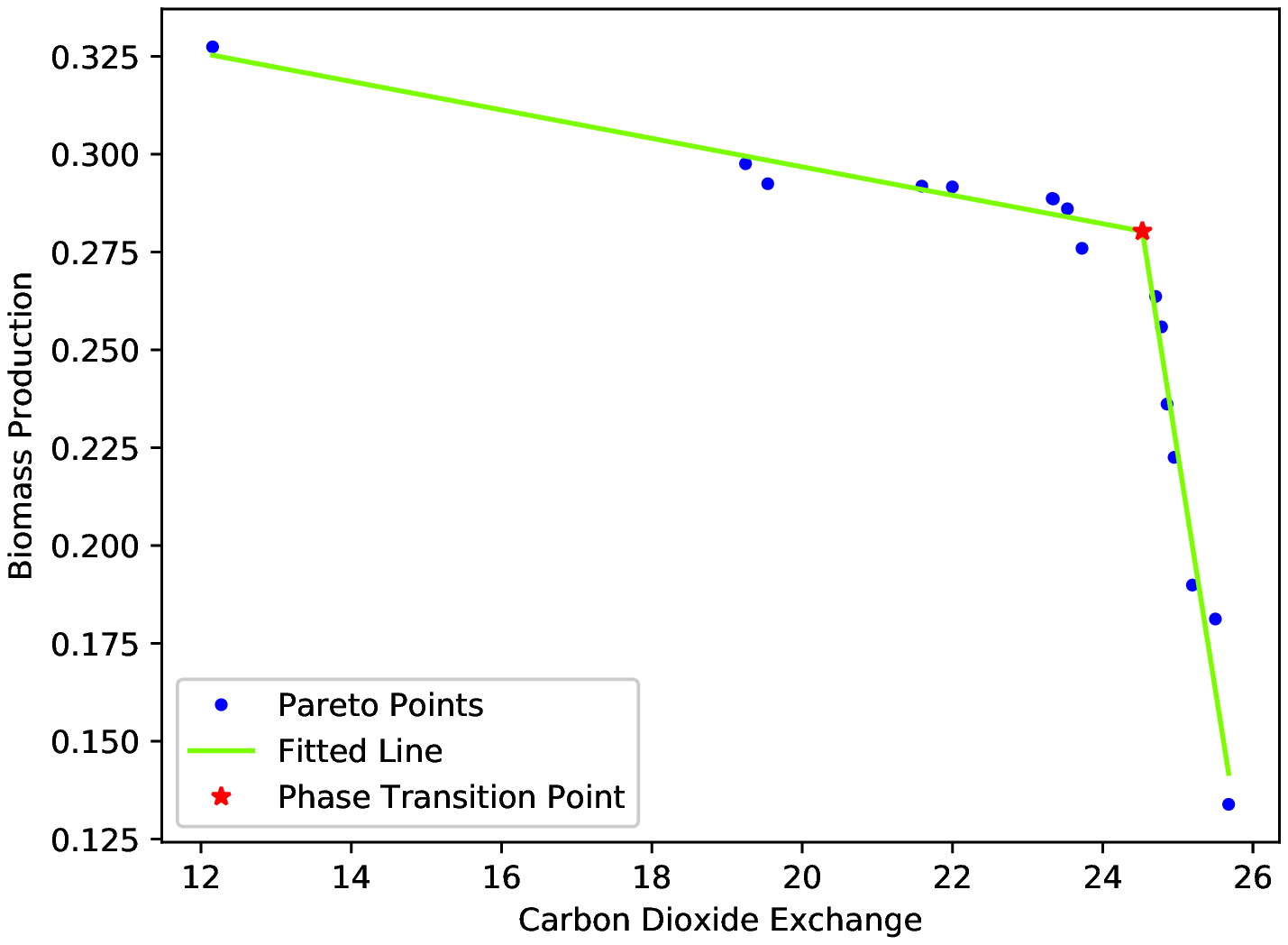}
	\caption{The Pareto fronts from the Oxygen and Carbon Dioxide trade-offs with Biomass. The Oxygen trade-off is particularly dense, whereas the Carbon Dioxide trade-off is very sparse.}
	\label{fig:init paretos}
\end{figure}

To analyse the performance of network regression with differing parameters, two test data sets will be used. One from the Oxygen, Bio\-mass trade-off shown at the beginning of section \ref{sec: res and disc}, but with the new flux bounds, the other from a Carbon Dioxide, Biomass trade-off. Figure \ref{fig:init paretos} shows the Pareto fronts in both cases. These were chosen as they are very different data sets, the Oxygen trade-off is very dense whereas the Carbon Dioxide trade-off has very few points.

\subsection{Pareto Reconstruction}
The regression parameters used were ($\alpha = 0.5, \lambda = 1.0, \tau = 99$). From this, the effectiveness of the network regression can be tested. Using exclusively the gene expression data, without knowledge of the metabolic or genetic network, the significant genes can be set to particular values, whilst leaving insignificant genes to take random values. The objectives can then be calculated using the generated gene expression data and compared against gene expression data comprised of pure noise. 

\begin{figure}[h]
	\centering
	\includegraphics[scale=0.25]{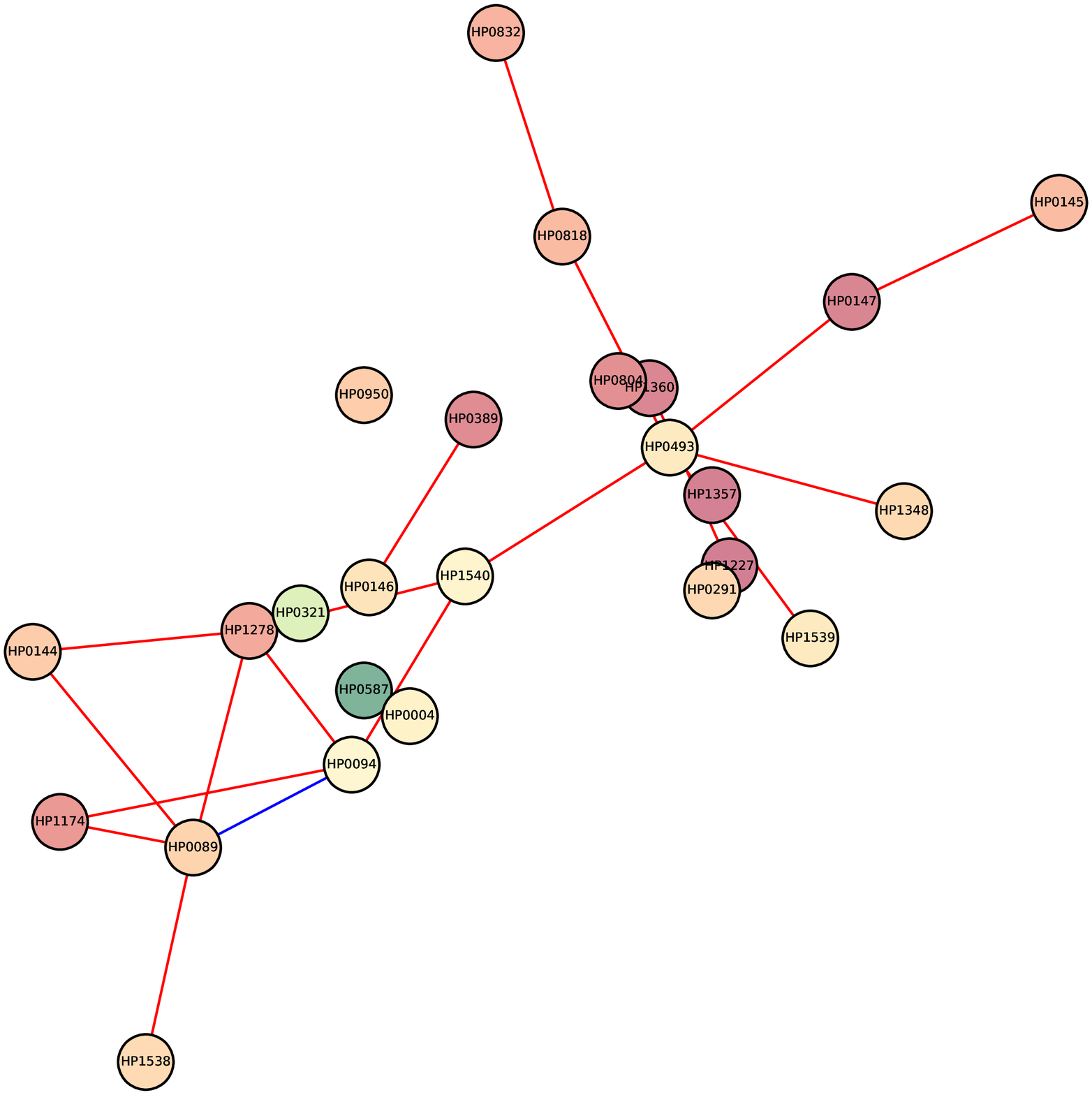}
	\includegraphics[scale=0.25]{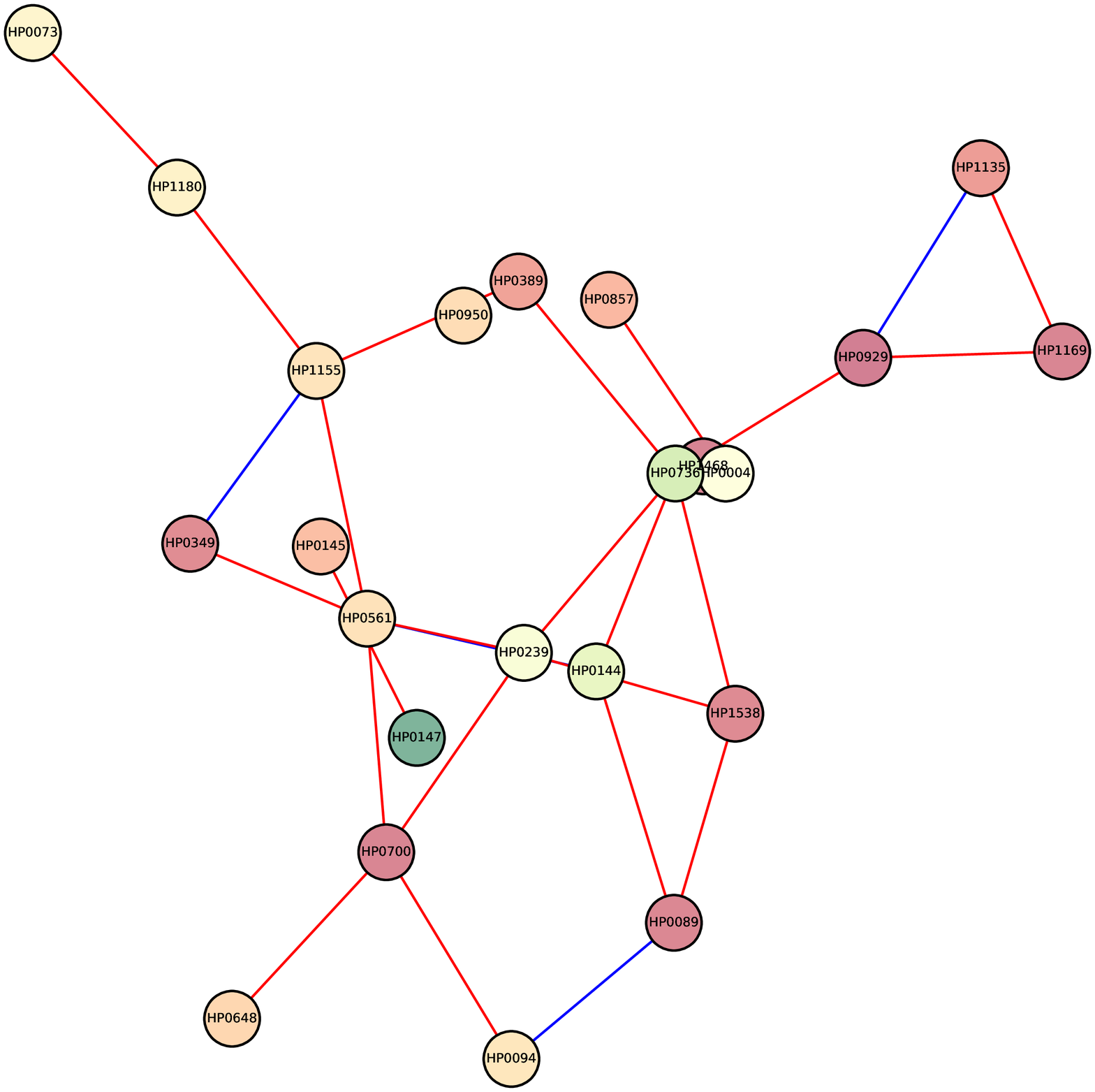}
    \caption{Significant gene networks for the Carbon Dioxide data. See figure \ref{fig:o2 significant genes} for further explanation.}
    \label{fig:co2 significant genes}
\end{figure}

\begin{figure}[h]
\centering
\includegraphics[scale=0.25]{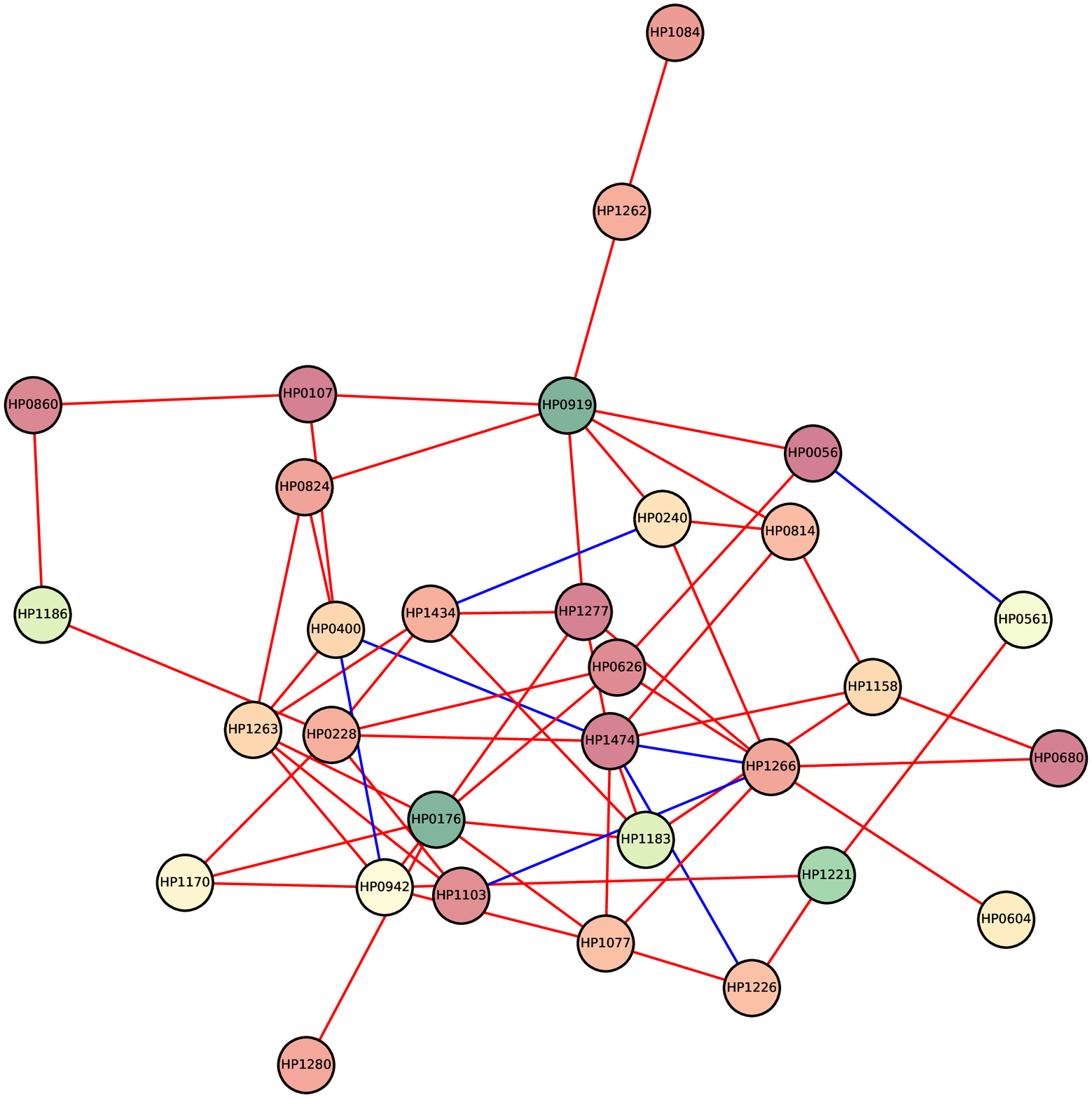}
	\includegraphics[scale=0.25]{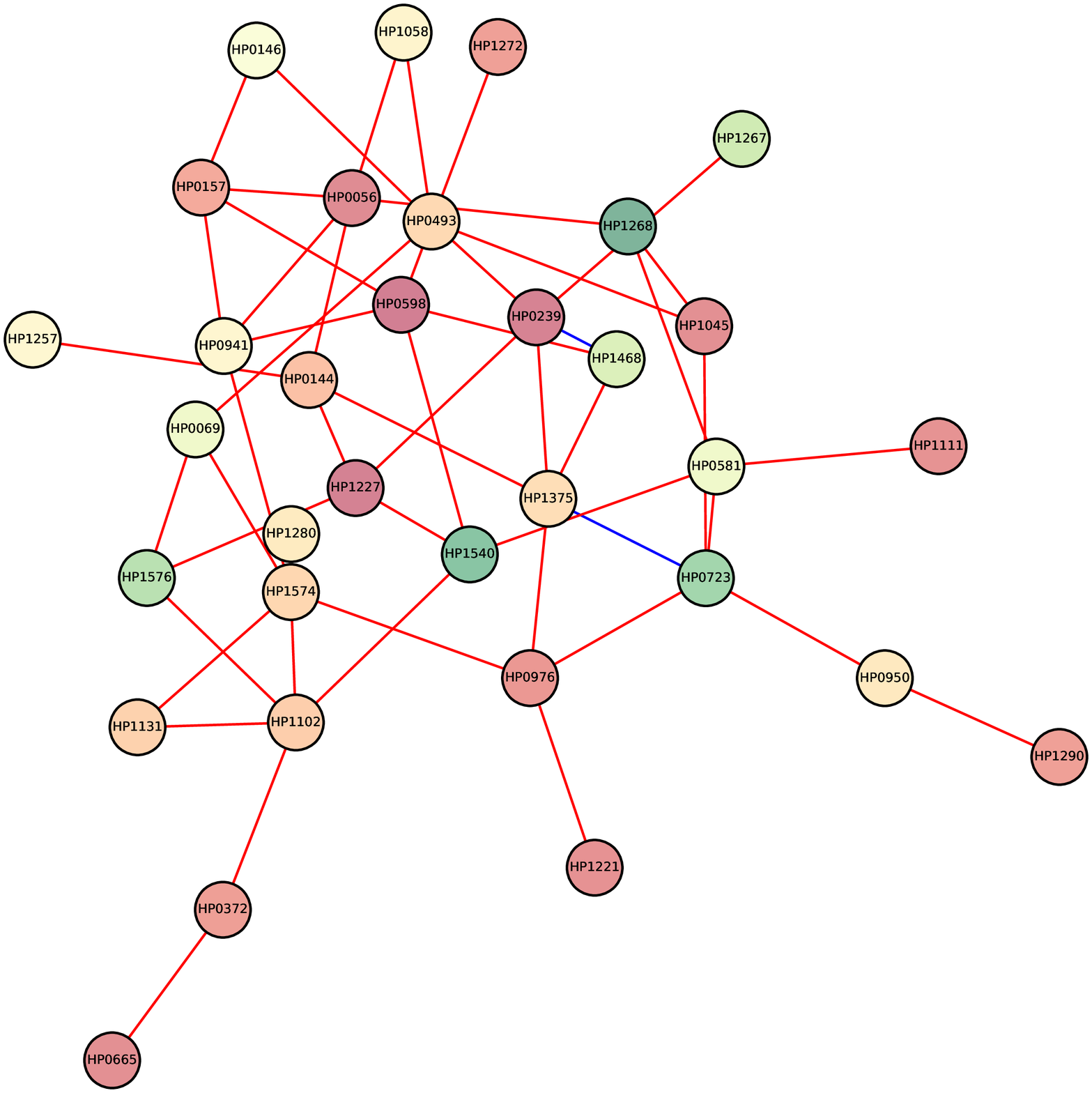}
	\caption{Significant gene networks for the Oxygen data. Networks for data on the left hand side of the Pareto front are above, and those for the right hand side are below. The colour of nodes indicates how significant the genes are, with green nodes being most significant and red the least significant. The blue edges are those contained within the network constraint, with $\tau = 99$, the red edges are those from the network formed with $\tau=90$.}
	\label{fig:o2 significant genes}
\end{figure}

The relevant measure of spread for the values of $\beta$ is the interquartile range. A significant value (outlier larger than the mean) is classed as a value one interquartile range larger than the mean. The network of significant genes is then the network formed from the subsection of the gene coexpression network containing only significant genes. These are shown in figures \ref{fig:co2 significant genes} and \ref{fig:o2 significant genes}. The green nodes are those that are more significant and red nodes are less significant. The names used are taken from the metabolic network. Red edges are those from a coexpression network with $\tau = 95$ compared to $99$ used in the regression itself which are indicated in blue. 

\begin{figure}[h]
	\centering
	\includegraphics[scale = 0.45]{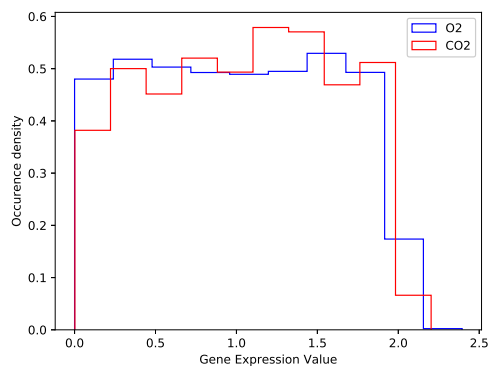}
	\caption{Histogram of the distribution of gene expression values in the Pareto datasets. This appears to be a uniform distribution between 0 and 2.}
	\label{fig:gene_express_dist}
\end{figure}

With the significant genes identified, a distribution of noise must be chosen for the insignificant genes. To choose this, we look at the distribution of gene expression values in the original dataset. From figure \ref{fig:gene_express_dist} the distribution  of insignificant genes appears to be uniform between 0 and 2.

Values for the significant genes, are chosen as follows. Key nodes are identified using the network reduction technique described in section \ref{ssec:network reduction}, values from the original data are selected for each of these nodes. Values for other significant nodes are chosen to be values from the original dataset corresponding to adjacent key nodes. By doing this the existing distributions of values is kept for each of the significant nodes and the network relations are enforced.

\begin{figure}[h]
	\centering
	\includegraphics[scale=0.43]{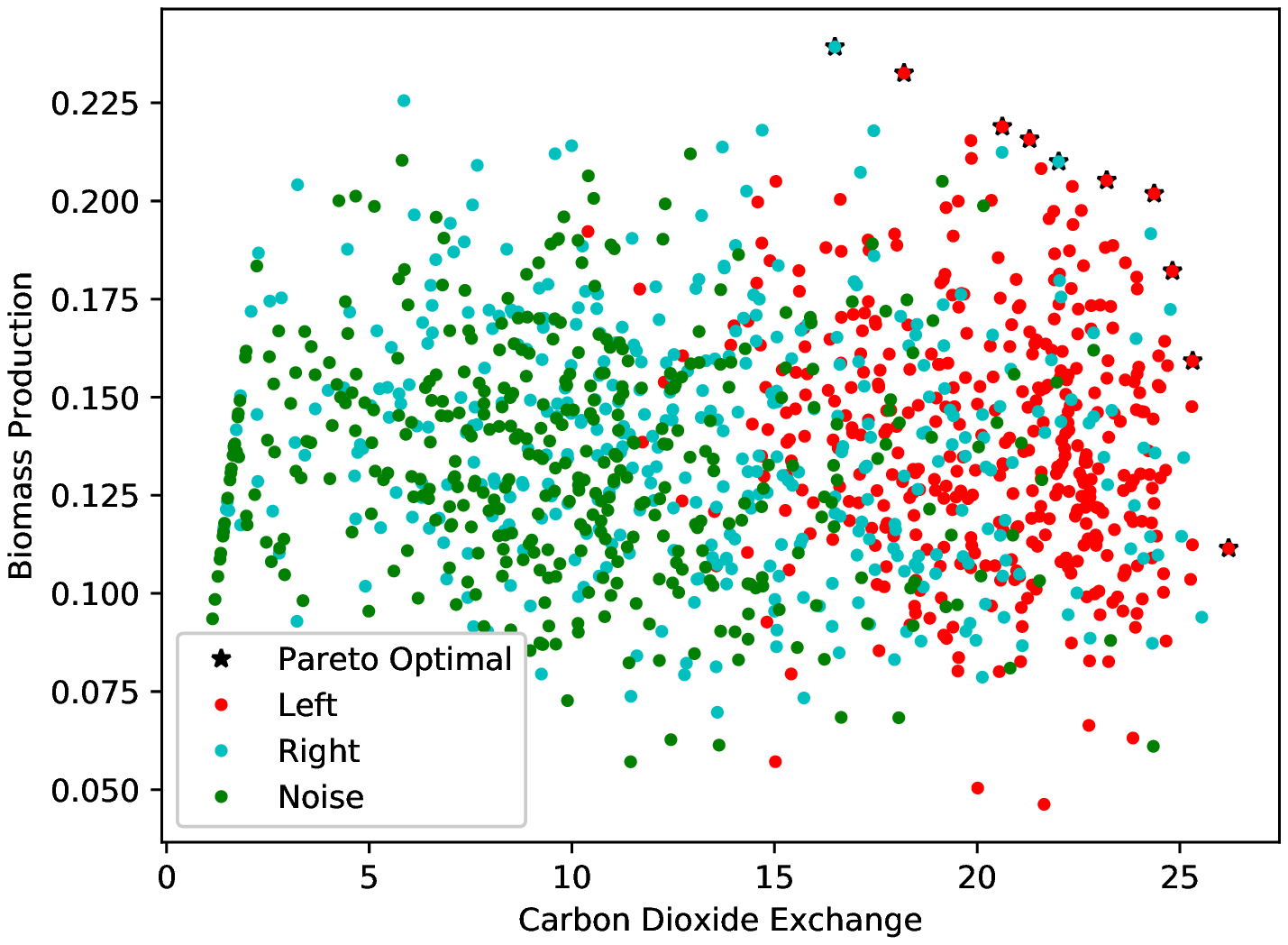}
	\includegraphics[scale=0.43]{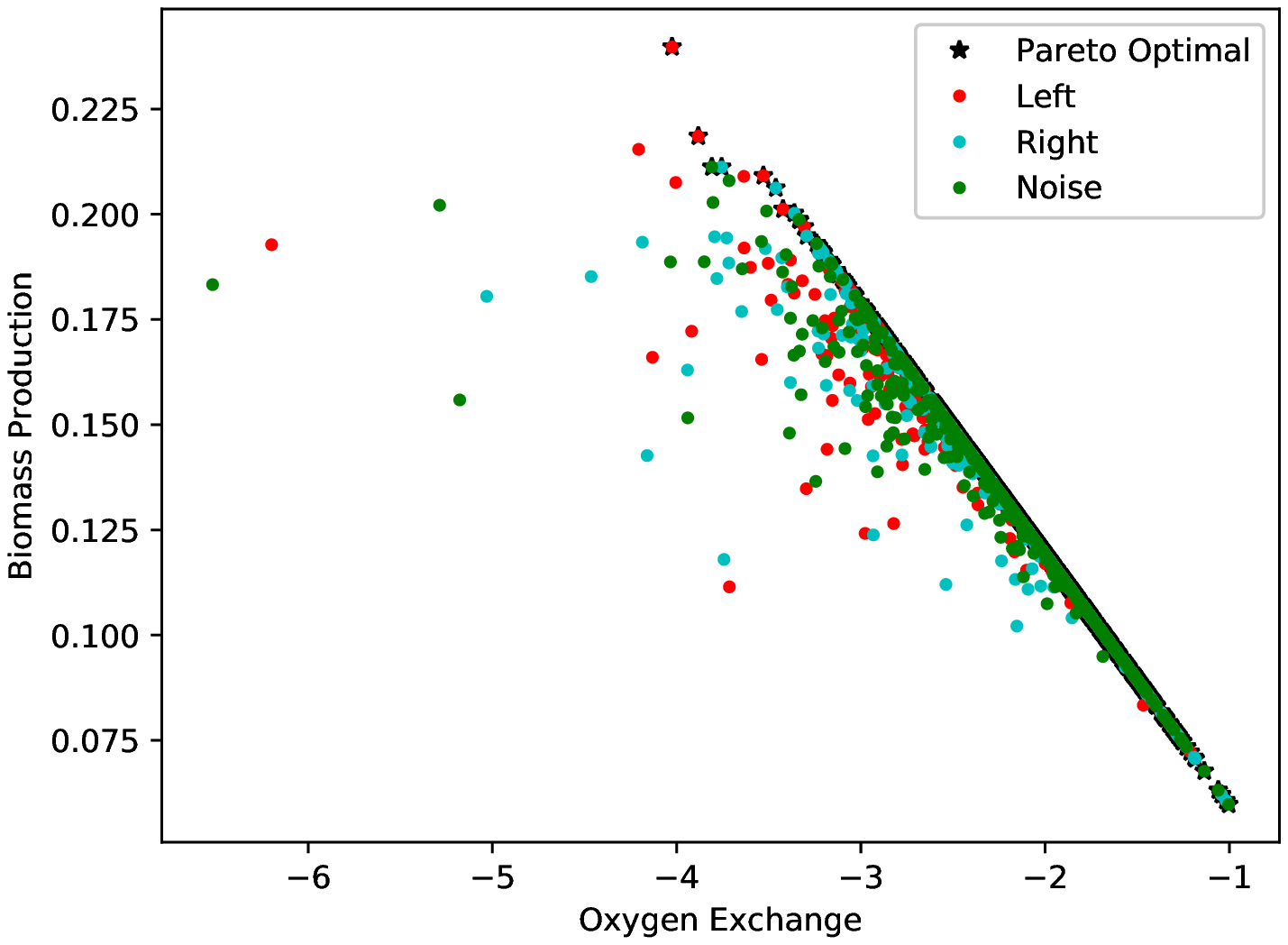}
	\caption{Reconstructed trade-offs using gene expression created based on the network regression. Points created from pure noise are in green, those from the regression on the left hand side of the Pareto data in red and right are blue. The Carbon Dioxide reconstruction appears to have worked well, whereas points based on Pareto data seem similar to those created from noise for the Oxygen trade-off.}
	\label{fig:reconstructions}	
\end{figure}

This process gives a new set of gene expression data, the objectives for this data can be calculated and a new Pareto front constructed. These reconstructions are shown in figure \ref{fig:reconstructions}, there are 800 points constructed from the regression (400 left, 400 right) and 400 points constructed from pure noise. The Oxygen reconstruction is not particularly successful, the random data appears to do as well at constructing the Pareto front as the reconstructed data. However, for points on the left hand side of the Pareto front, the reconstructed data is a more successful. The Carbon Dioxide reconstruction is, however, clearly more successful. The red points constructed from the regression on the left hand portion of the Pareto data are considerably more successful than the green noise points.

\section{Conclusions}
\label{sec:conc}
	
Networked Cox regression was used, for the first time, to analyse data from a multi-omic metabolic model of \textit{H. Pylori}. It is clear that the flux bounds used within the metabolic model are significant, this prompted the use of loopless Flux Balance Analysis to choose the flux bounds in this instance. 

Two example datasets were then chosen to tune the parameters for the network regression, a Biomass and Oxygen trade-off and a Biomass and Carbon Dioxide trade-off. These were chosen as the Carbon Dioxide trade-off has a small number of points whereas the Oxygen trade-off has a far larger number of points. The regression parameters used were ($\alpha = 0.5, \lambda = 1.0, \tau = 99$).

With regression variables for each dataset chosen, an attempt was made to reconstruct the original Pareto data. The distribution of gene expression values in the original datasets were used to choose a noise distribution, uniform from $0$ to $2$. Significant genes were then chosen from the regression variables, those that were one interquartile range above the mean. The reduced network of these genes was then constructed. Significant genes in the reduced network selected values from the original dataset, nodes adjacent to the significant nodes selected values corresponding to those of the adjacent significant nodes. Non-significant genes were left as noise. From this new gene expression datasets were generated and the objectives calculated. The reconstruction for the Oxygen data did not seem to be very different from the noise distribution, except for the values on the left hand side of the Pareto front. In contrast, the reconstruction of the Carbon Dioxide data was successful, with exclusively reconstructed points forming the Pareto front.

In summary we present a computational framework for the integration and analysis of biological network data. Such a challenge can be mapped into three main computational problems:
(a) data integration: merging data at different scales (multi-omic data) is necessary to understand the different levels of network evolution and how these levels interact each other;
(b) high dimensionality: multi-omic data exists in a high dimensional matrix describing the expression level of each gene or protein;
(c) merging statistical and biological knowledge in genomic data analysis: accurate data analysis cannot be performed using only statistical approaches, but rather \textit{a priori} biological knowledge needs to be included in the final computational framework. Indeed, appropriate data investigation should be based both on statistical and biological knowledge in order to merge the information that can be extracted from the data (statistical information) and the biological knowledge
(marker related information already known in literature).

The methodology and the software presented in this work could be used for a variety of research useful to the molecular biology community. For example, in bacterial genomes there are high levels of functional redundancies which represent mechanisms employed in cells to achieve robustness. These mechanisms allow, under different environmental conditions, very different sets of reactions to compensate for one another \cite{Sambamoorthy2018}.
An interesting application of this method is the possibility of integrating different metabolic networks or different multi omic networks. Another application is the study of the 
metabolic model of the human (for example from Recon 2) and reconstructions for the
dozens of bacterial species forming the gut microbiome. 
	
\onecolumn
\bibliographystyle{spphys}
\bibliography{final_report_bib}

\appendix
\section{Code Guide}

The code is written in such a way as to enable the user to avoid as much of the underlying complexity as possible. A guide to producing results similar to those listed in this paper follows;

\begin{lstlisting}[language=Python,name=gen]
import cobra
import pareto
import pa_re
import ne_re
\end{lstlisting}

Cobra, the pareto, pareto reconstruction and network regression files are imported. Next the model is loaded and the objectives are chosen. The example listed is for the \textit{E. coli} model found on BiGG models.

\begin{lstlisting}[language=Python,name=gen]
model_str = 'iJO1366.json'
model = cobra.io.load_json_model(model_str)
obj1_str = ' BIOMASS_Ec_iJO1366_core_53p95M'
obj1 = model.reactions.get_by_id(obj1_str).flux_expression
obj2_str = 'EX_O2_e'
obj2 = model.reactions.get_by_id(obj2_str).flux_expression
filename = 'test_data.json'
\end{lstlisting}
Basic parameters for the dataset generation, regression and reconstruction can then be set.
\begin{lstlisting}[language=Python,name=gen]
LAMBD = 1.0
ALPHA = 0.5
CUTOFF = 99
GENS = 20
INDIV = 200
NODES = 40
RECON_POINTS = 200
CORES = 0
\end{lstlisting}
Here, the values of ($\lambda = 1.0, \alpha = 0.5, \tau = 0.99$) are used and typical values for the number of generations, individuals per generation and reconstruction points. This would use the single threaded version of the code, but `CORES' can be set to the desired number of cores to be used in order to enable multithreading (with 0 using the sequential version).

Running Pareto data generation and storing to initial JSON file.

The initial Pareto data can then be generated. This data is converted into a dictionary which can then be easily stored within a JSON file. 
\begin{lstlisting}[language=Python,name=gen]
pops, vals, pareto_data = pareto.pareto(GENS, INDIV, model, obj1, obj2,cores = CORES)
to_save = {'obj1_str': str(obj1), 'obj2_str': str(obj2), 'model' : model_str,'pareto': [{'obj1': p.fitness.values[0], 'obj2':p.fitness.values[1],'gene_set': list(p)} for p in pareto_data]}
with open(filename, 'w') as outfile:
	json.dump(to_save, outfile)
\end{lstlisting}

Running the network regression and storing in JSON file is simply done with a single command.
\begin{lstlisting}[language=Python,name=gen]
ne_re.add_linear_regression(filename, CUTOFF)
\end{lstlisting}

The Pareto Reconstruction can then be added to the JSON file.
\begin{lstlisting}[language=Python,name=gen]
pareto_left,pareto_right,pareto_noise,pareto_y,pareto_x = pa_re.reconstruct(filename, NODES, RECON_POINTS, model, obj1, obj2,cores=CORES)
\end{lstlisting}

These individual parts are all stored within the JSON file, so can then be easily plotted at a later date, and a batch script can easily be used to generate large amounts of complete data.
\end{document}